\pdfoutput=1

\documentclass[12pt,a4paper]{article}

\usepackage{ifthen} 
\newboolean{pdflatex}
\setboolean{pdflatex}{true} 

\newboolean{articletitles}
\setboolean{articletitles}{true} 

\newboolean{uprightparticles}
\setboolean{uprightparticles}{false} 

\newboolean{inbibliography}
\setboolean{inbibliography}{false} 

\def\paperauthors{LHCb collaboration} 
\def\paperasciititle{Model-independent observation of exotic contributions to B0->JpsiK+pi- decays} 
\def\papertitle{Model-independent observation of exotic contributions to \BdToJPsiKpi decays} 
\def\paperkeywords{{High Energy Physics}, {LHCb}} 
\def\papercopyright{\the\year\ CERN for the benefit of the LHCb collaboration} 
\def\paperlicence{CC-BY-4.0 licence}
\def\paperlicenceurl{https://creativecommons.org/licenses/by/4.0/}

\newboolean{isPRL}
\setboolean{isPRL}{false}

\usepackage[top=1in, bottom=1.25in, left=1in, right=1in]{geometry}

\usepackage{microtype}
\usepackage{lineno}  
\usepackage{xspace} 
\usepackage{caption} 

\usepackage{graphicx}  
\usepackage{color}
\usepackage{colortbl}
\graphicspath{{./figs/}} 
\DeclareGraphicsExtensions{.pdf,.PDF,png,.PNG}

\usepackage{amsmath} 
\usepackage{amssymb}
\usepackage{amsfonts}
\usepackage{upgreek} 

\newcommand*\patchAmsMathEnvironmentForLineno[1]{%
\expandafter\let\csname old#1\expandafter\endcsname\csname #1\endcsname
\expandafter\let\csname oldend#1\expandafter\endcsname\csname
end#1\endcsname
 \renewenvironment{#1}%
   {\linenomath\csname old#1\endcsname}%
   {\csname oldend#1\endcsname\endlinenomath}%
}
\newcommand*\patchBothAmsMathEnvironmentsForLineno[1]{%
  \patchAmsMathEnvironmentForLineno{#1}%
  \patchAmsMathEnvironmentForLineno{#1*}%
}
\AtBeginDocument{%
\patchBothAmsMathEnvironmentsForLineno{equation}%
\patchBothAmsMathEnvironmentsForLineno{align}%
\patchBothAmsMathEnvironmentsForLineno{flalign}%
\patchBothAmsMathEnvironmentsForLineno{alignat}%
\patchBothAmsMathEnvironmentsForLineno{gather}%
\patchBothAmsMathEnvironmentsForLineno{multline}%
\patchBothAmsMathEnvironmentsForLineno{eqnarray}%
}

\usepackage{hyperxmp}

\usepackage[pdftex,
            pdfauthor={\paperauthors},
            pdftitle={\paperasciititle},
            pdfkeywords={\paperkeywords},
            pdfcopyright={Copyright (C) \papercopyright},
            pdflicenseurl={\paperlicenceurl}]{hyperref}

\usepackage[colorinlistoftodos,textsize=scriptsize]{todonotes}

\usepackage[all]{hypcap} 

\usepackage{xspace} 
\usepackage{upgreek}

\def\lhcb   {\mbox{LHCb}\xspace}

\def\MagUp {\mbox{\em Mag\kern -0.05em Up}\xspace}

\ifthenelse{\boolean{uprightparticles}}%
{

 \def\Pmu         {\ensuremath{\upmu}\xspace}

 \def\Ppi         {\ensuremath{\uppi}\xspace}

 \def\Ppsi        {\ensuremath{\uppsi}\xspace}

 \def\PDelta      {\ensuremath{\Delta}\xspace}                 
 \def\PXi         {\ensuremath{\Xi}\xspace}                 
 \def\PLambda     {\ensuremath{\Lambda}\xspace}                 
 \def\PSigma      {\ensuremath{\Sigma}\xspace}                 
 \def\POmega      {\ensuremath{\Omega}\xspace}                 
 \def\PUpsilon    {\ensuremath{\Upsilon}\xspace}

 \def\PB      {\ensuremath{\mathrm{B}}\xspace}                 
                  
 \def\PD      {\ensuremath{\mathrm{D}}\xspace}

 \def\PJ      {\ensuremath{\mathrm{J}}\xspace}                 
 \def\PK      {\ensuremath{\mathrm{K}}\xspace}

 \def\Pb      {\ensuremath{\mathrm{b}}\xspace}

 \def\Pi      {\ensuremath{\mathrm{i}}\xspace}

 \def\thebaroffset{0.0em}
}
{

 \def\Pmu         {\ensuremath{\mu}\xspace}

 \def\Ppi         {\ensuremath{\pi}\xspace}

 \def\Ppsi        {\ensuremath{\psi}\xspace}                 
                  
 \mathchardef\PDelta="7101
 \mathchardef\PXi="7104
 \mathchardef\PLambda="7103
 \mathchardef\PSigma="7106
 \mathchardef\POmega="710A
 \mathchardef\PUpsilon="7107
                  
 \def\PB      {\ensuremath{B}\xspace}                 
                  
 \def\PD      {\ensuremath{D}\xspace}

 \def\PJ      {\ensuremath{J}\xspace}                 
 \def\PK      {\ensuremath{K}\xspace}

 \def\Pb      {\ensuremath{b}\xspace}

 \def\Pi      {\ensuremath{i}\xspace}

 \def\thebaroffset{0.18em}
}
\newcommand{\offsetoverline}[2][\thebaroffset]{\kern #1\overline{\kern -#1 #2}}%

\makeatletter
\ifcase \@ptsize \relax
  \newcommand{\miniscule}{\@setfontsize\miniscule{4}{5}}
\or
  \newcommand{\miniscule}{\@setfontsize\miniscule{5}{6}}
\or
  \newcommand{\miniscule}{\@setfontsize\miniscule{5}{6}}
\fi
\makeatother

\DeclareRobustCommand{\optbar}[1]{\shortstack{{\miniscule (\rule[.5ex]{1.25em}{.18mm})}
  \\ [-.7ex] $#1$}}

\def\mup        {{\ensuremath{\Pmu^+}}\xspace}
\def\mun        {{\ensuremath{\Pmu^-}}\xspace} 

\def\bquark    {{\ensuremath{\Pb}}\xspace}

\def\pion   {{\ensuremath{\Ppi}}\xspace}

\def\pip    {{\ensuremath{\pion^+}}\xspace}
\def\pim    {{\ensuremath{\pion^-}}\xspace}

\def\kaon    {{\ensuremath{\PK}}\xspace}

\def\KorKbar {\kern \thebaroffset\optbar{\kern -\thebaroffset \PK}{}\xspace}

\def\Kp      {{\ensuremath{\kaon^+}}\xspace}
\def\Km      {{\ensuremath{\kaon^-}}\xspace}

\def\DorDbar {\kern \thebaroffset\optbar{\kern -\thebaroffset \PD}\xspace}

\def\B       {{\ensuremath{\PB}}\xspace}
\def\Bbar    {{\ensuremath{\offsetoverline{\PB}}}\xspace}

\def\BorBbar {\kern \thebaroffset\optbar{\kern -\thebaroffset \PB}\xspace}
\def\Bz      {{\ensuremath{\B^0}}\xspace}
\def\Bzb     {{\ensuremath{\Bbar{}^0}}\xspace}

\def\Bd      {{\ensuremath{\B^0}}\xspace}

\def\jpsi     {{\ensuremath{{\PJ\mskip -3mu/\mskip -2mu\Ppsi\mskip 2mu}}}\xspace}
\def\psitwos  {{\ensuremath{\Ppsi{(2S)}}}\xspace}

\def\Y#1S{\ensuremath{\PUpsilon{(#1S)}}\xspace}

\def\Lz          {{\ensuremath{\PLambda}}\xspace}

\def\LorLbar     {\kern \thebaroffset\optbar{\kern -\thebaroffset \PLambda}\xspace}

\def\Lb           {{\ensuremath{\Lz^0_\bquark}}\xspace}

\def\to                 {\ensuremath{\rightarrow}\xspace}

\def\qsq       {{\ensuremath{q^2}}\xspace}

\def\CP                {{\ensuremath{C\!P}}\xspace}

\def\AT#1     {\ensuremath{A_{\mathrm{T}}^{#1}}\xspace}           

\def\ctl       {\ensuremath{\cos{\theta_\ell}}\xspace}

\def\C#1      {\ensuremath{\mathcal{C}_{#1}}\xspace}                       
\def\Cp#1     {\ensuremath{\mathcal{C}_{#1}^{'}}\xspace}                    
\def\Ceff#1   {\ensuremath{\mathcal{C}_{#1}^{\mathrm{(eff)}}}\xspace}        
\def\Cpeff#1  {\ensuremath{\mathcal{C}_{#1}^{'\mathrm{(eff)}}}\xspace}       
\def\Ope#1    {\ensuremath{\mathcal{O}_{#1}}\xspace}                       
\def\Opep#1   {\ensuremath{\mathcal{O}_{#1}^{'}}\xspace}                    

\newcommand{\aunit}[1]{\ensuremath{\text{\,#1}}}       

\newcommand{\tev}{\aunit{Te\kern -0.1em V}\xspace}
\newcommand{\gev}{\aunit{Ge\kern -0.1em V}\xspace}
\newcommand{\mev}{\aunit{Me\kern -0.1em V}\xspace}
\newcommand{\kev}{\aunit{ke\kern -0.1em V}\xspace}
\newcommand{\ev}{\aunit{e\kern -0.1em V}\xspace}
\newcommand{\mevc}{\ensuremath{\aunit{Me\kern -0.1em V\!/}c}\xspace}
\newcommand{\gevc}{\ensuremath{\aunit{Ge\kern -0.1em V\!/}c}\xspace}
\newcommand{\mevcc}{\ensuremath{\aunit{Me\kern -0.1em V\!/}c^2}\xspace}
\newcommand{\gevcc}{\ensuremath{\aunit{Ge\kern -0.1em V\!/}c^2}\xspace}

\def\fb   {\ensuremath{\aunit{fb}}\xspace}
\def\invfb   {\ensuremath{\fb^{-1}}\xspace}

\def\deriv {\ensuremath{\mathrm{d}}}

\def\gsim{{~\raise.15em\hbox{$>$}\kern-.85em
          \lower.35em\hbox{$\sim$}~}\xspace}
\def\lsim{{~\raise.15em\hbox{$<$}\kern-.85em
          \lower.35em\hbox{$\sim$}~}\xspace}

\def\evtgen     {\mbox{\textsc{EvtGen}}\xspace}

\def\photos     {\mbox{\textsc{Photos}}\xspace}

\def\pythia     {\mbox{\textsc{Pythia}}\xspace}

\xspace

\def\tell1  {TELL1\xspace}
\def\ukl1   {UKL1\xspace}

\newcommand{\ie}{\mbox{\itshape i.e.}\xspace}

\def\BdToJPsiKpi  {\ensuremath{\Bd\to\jpsi\Kp\pim}}
\def\BdToJPsimmKpi {\ensuremath{\Bd\to\jpsi(\to \mup \mun)\Kp\pim} }
\def\mkpi  {\ensuremath{m(\Kp\pim)}\xspace}
\def\kpi{\ensuremath{\Kp\pim}\xspace}
\def\mJkpi{\ensuremath{m(\jpsi\Kp\pim)}\xspace}
\def\thetal {\ensuremath{\theta_l}\xspace}
\def\thetav {\ensuremath{\theta_V}\xspace}
\def\ctv {\ensuremath{\cos \theta_V}\xspace}
\def\ctl {\ensuremath{\cos \theta_\ell}\xspace}
\def\nmaxk  {\ensuremath{n^k_{\rm max}}\xspace}
\def\jmaxk  {\ensuremath{J^k_{\rm max}}\xspace}

\newcommand {\img} {\, Im}
\newcommand {\rel} {\, Re}

\def\hzsq        {\ensuremath{|H_0|^2}\xspace}

\def\ssq         {\ensuremath{|S|^2}\xspace}
\def\dzsq        {\ensuremath{|D_0|^2}\xspace}

\def\hpasq        {\ensuremath{|H_\parallel|^2}\xspace}
\def\hpesq        {\ensuremath{|H_\perp|^2}\xspace}
\def\dpasq        {\ensuremath{|D_\parallel|^2}\xspace}
\def\dpesq        {\ensuremath{|D_\perp|^2}\xspace}

\def\rhzdz       {\ensuremath{Re(H_0 D^{\ast}_0)}\xspace}

\def\rshz        {\ensuremath{Re(S H_0^{ \ast})}\xspace}

\def\rsdz        {\ensuremath{Re(S D_0^{ \ast})}\xspace}

\usepackage{cite} 
\usepackage{mciteplus}
\usepackage{subfigure}
\usepackage{bm}

\usepackage{longtable} 

\begin{document}

\renewcommand{\thefootnote}{\fnsymbol{footnote}}
\setcounter{footnote}{1}


\begin{titlepage}
\pagenumbering{roman}

\vspace*{-1.5cm}
\centerline{\large EUROPEAN ORGANIZATION FOR NUCLEAR RESEARCH (CERN)}
\vspace*{1.5cm}
\noindent
\begin{tabular*}{\linewidth}{lc@{\extracolsep{\fill}}r@{\extracolsep{0pt}}}
\ifthenelse{\boolean{pdflatex}}
{\vspace*{-1.5cm}\mbox{\!\!\!\includegraphics[width=.14\textwidth]{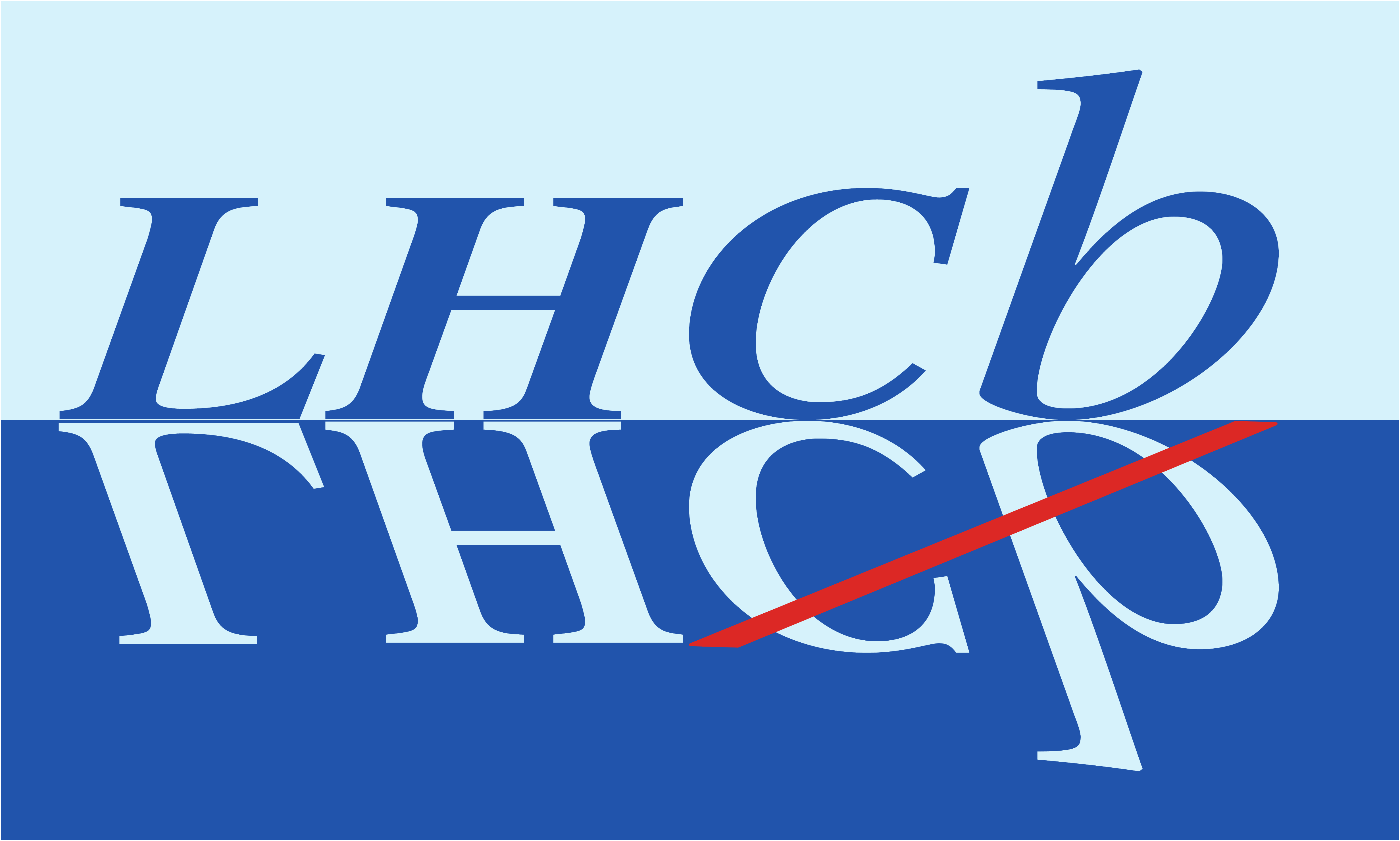}} & &}%
{\vspace*{-1.2cm}\mbox{\!\!\!\includegraphics[width=.12\textwidth]{lhcb-logo.eps}} & &}%
\\
 & & CERN-EP-2018-330 \\  
 & & LHCb-PAPER-2018-043 \\  
 & & \today \\ 
 & & \\
\end{tabular*}

\vspace*{4.0cm}

{\normalfont\bfseries\boldmath\huge
\begin{center}
  \papertitle 
\end{center}
}

\vspace*{2.0cm}

\begin{center}
\paperauthors\footnote{Authors are listed at the end of this paper.}
\end{center}

\vspace{\fill}

\begin{abstract}
  \noindent An angular analysis of $\BdToJPsiKpi$ decays is performed, using proton-proton collision data corresponding to an integrated luminosity of 3\invfb collected with the LHCb detector. The \mkpi spectrum is divided into fine bins. In each \mkpi bin, the hypothesis that the three-dimensional angular distribution can be described by structures induced only by $K^\ast$ resonances is examined, making minimal assumptions about the $K^+\pi^-$ system. The data reject the $K^\ast$-only hypothesis with a large significance, implying the observation of exotic contributions in a model-independent fashion. Inspection of the $m(\jpsi \pim)$ versus $\mkpi$ plane suggests structures near $m(\jpsi \pim)=4200$~\mev and 4600~\mev.

\end{abstract}

\vspace*{2.0cm}

\begin{center}
  Published in Phys. Rev. Lett. 122, 152002 (2019)
\end{center}

\vspace{\fill}

{\footnotesize 
\centerline{\copyright~\papercopyright. \href{\paperlicenceurl}{\paperlicence}.}}
\vspace*{2mm}

\end{titlepage}


\newpage
\setcounter{page}{2}
\mbox{~}
%
%
%
%

\cleardoublepage


\renewcommand{\thefootnote}{\arabic{footnote}}
\setcounter{footnote}{0}



\pagestyle{plain} 
\setcounter{page}{1}
\pagenumbering{arabic}


%

In the Standard Model, the quark model allows for hadrons comprising any number of valence quarks, as long as they are colour-singlet states. Yet, after decades of searches, the reason why the vast majority of hadrons are built out of only quark-antiquark (meson) or three-quark (baryon) combinations remains a mystery. The best known exception is the $Z(4430)^-$ resonance with spin-parity $1^-$ and width $\Gamma=172\pm13$\mev~\cite{Aaij:2014jqa,Choi:2007wga},\footnote{Natural units with $\hslash=c=1$ are used throughout the document.} which has minimal quark content $c\bar{c}\bar{u}\bar{d}$, and is therefore manifestly exotic, \ie, has components that are neither quark-antiquark or three-quark combinations. The only confirmed decay of the $Z(4430)^-$ state is via ${Z(4430)^-\to \psitwos \pi^-}$, as seen in ${B^0\to \psitwos K^+\pi^-}$ decays~\cite{Aaij:2014jqa}.\footnote{The inclusion of charge-conjugate decay modes is implied throughout.} The corresponding ${Z(4430)^-\to \jpsi \pi^-}$ decay rate is suppressed by at least a factor of ten~\cite{Chilikin:2014bkk}. The authors of Ref.~\cite{Brodsky:2014xia} surmise that in a dynamical diquark picture, this is because of a larger overlap of the $Z(4430)^-$ radial wavefunction with the excited state \psitwos than with the ground state \jpsi. For the $B^0\to \jpsi K^+\pi^-$ channel, the Belle collaboration~\cite{Chilikin:2014bkk} has reported the observation of a new exotic $Z(4200)^-$ resonance decaying to $\jpsi \pi^-$, that might correspond to the structure in $m(\psitwos \pim)$ seen in Ref.~\cite{Aaij:2014jqa} at around the same mass.

\begin{figure}
 \includegraphics[width=\linewidth]{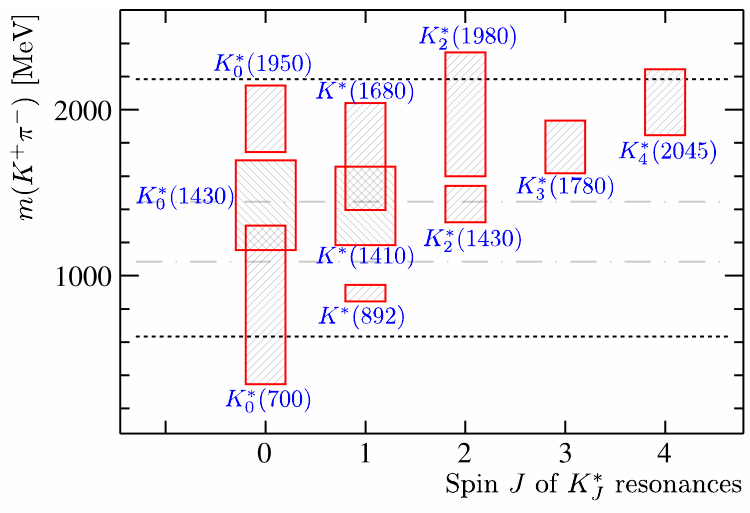}
 \caption{Spectrum of $K^\ast_J$ resonances from Ref.~\cite{PDG2018}, with the vertical span of the boxes indicating $\pm\Gamma_0$, where $\Gamma_0$ is the width of each resonance. The horizontal dashed lines mark the \mkpi physical region for ${B^0\to \jpsi K^+\pi^-}$ decays, while the dot-dashed lines mark the specific region, ${\mkpi\in [1085,1445]}$~\mev, employed for determining the significance of exotic contributions.} 
\label{fig:Kst_res}
\end{figure}

A generic concern in searches for broad exotic states like the $Z(4430)^-$ resonance is disentangling contributions from non-exotic components. For $B^0\to \psi^{(')} K^+\pi^-$ decays,\footnote{Here $\psi$ denotes the ground state $\jpsi$ and $\psi'$ denotes the excited state \psitwos.} the latter comprise different $K^\ast_J$ resonances with spin $J$, that decay to $K^+\pi^-$. Figure~\ref{fig:Kst_res} shows the $K^\ast_J$ spectrum, which has multiple, overlapping, and poorly measured states. The bulk of the measurements come from the LASS $K^+\pi^-$ scattering experiment~\cite{Aston:1987ir}. In particular, the decay $B^0\to \jpsi K^+\pim$ is known to be dominated by $K^\ast_J$ resonances, with an exotic fit fraction of only $2.4\%$~\cite{Chilikin:2014bkk}, compared to a $10.3\%$ contribution from the $Z(4430)^-$ for $B^0\to \psitwos K^+\pim$~\cite{Chilikin:2013tch}. This smaller exotic fit fraction for the \jpsi case makes it pertinent to study the evidence of exotic contributions in a manner independent of the dominant but poorly understood $K^\ast_J$ spectrum. 

The BaBar collaboration~\cite{Aubert:2008aa} has performed a model-independent analysis of ${B^0\to \psi^{(')} K^+\pi^-}$ decays making minimal assumptions about the $K^\ast_J$ spectrum, using two-dimensional~(2D) moments in the variables \mkpi and the $K^+$ helicity angle, \thetav. The key feature of this  approach is that no information on the exact content of the $K^\ast_J$ states, including their masses, widths and \mkpi-dependent lineshapes, is required. An amplitude analysis would require the accurate description of the $K^\ast_J$ lineshapes which depend on the underlying production dynamics. The model-independent procedure bypasses these problems, requiring only knowledge of the highest spin, $J_{\rm max}$, among all the contributing $K^\ast_J$ states, for a given \mkpi bin. Within uncertainties, the $m(\jpsi\pi^-)$ spectrum in the BaBar data was found to be adequately described using just $K^\ast_J$ states, without the need for exotic contributions.

In this Letter, a four-dimensional~(4D) angular analysis of $B^0\to \jpsi K^+\pi^-$ decays with $\jpsi\to \mup\mun$ is reported, employing the Run~1 LHCb dataset. The data sample corresponds to a signal yield approximately $40$ and $20$ times larger than those of the corresponding BaBar~\cite{Aubert:2008aa} and Belle~\cite{Chilikin:2013tch} analyses, respectively. The larger sample size allows analysis of the differential rate as a function of the four variables, \mkpi, $\thetav$, $\thetal$ and $\chi$, that fully describe the decay topology. The lepton helicity angle, $\thetal$, and the azimuthal angle, $\chi$, between the $(\mup\mun)$ and $(K^+\pi^-)$ decay planes, were integrated over in the BaBar 2D analysis~\cite{Aubert:2008aa}. The present 4D analysis therefore benefits from a significantly better sensitivity to exotic components than the previous 2D analysis.

The LHCb detector is a single-arm forward spectrometer covering the pseudorapidity range $2 < \eta < 5$ and is described in detail in Ref.~\cite{Alves:2008zz}. Samples of simulated events are used to obtain the detector efficiency and optimise the selection. The $pp$ collisions are generated using \pythia~\cite{Sjostrand:2006za,*Sjostrand:2007gs} with a specific \lhcb configuration~\cite{LHCb-PROC-2010-056}. Decays of hadronic particles are described by \evtgen~\cite{Lange:2001uf}, in which final-state radiation is generated using \photos~\cite{Golonka:2005pn}. Dedicated control samples are employed to calibrate the simulation for agreement with the data.

The selection procedure is the same as in Refs.~\cite{Aaij:2015oid,Aaij:2016kqt} for the rare decay ${B^0 \to \mu^+ \mu^- K^+\pi^-}$, with the additional requirement that the $m(\mu^+\mu^-)$ mass is constrained to the known \jpsi mass via a kinematic fit~\cite{Hulsbergen:2005pu}. The data sample is divided into 35 fine bins in $\mkpi$ such that the $\mkpi$-dependence can be neglected inside a given bin, and each subsample is processed independently. The bin-widths vary depending on the data sample size in a given \mkpi region. Backgrounds from $B^+\to \jpsi K^+$, $B^0_s\to \jpsi K^+K^-$ and $\Lb\to \jpsi p K^-$ decays are reduced to a level below $1\%$ of the signal yield at the selection stage using the excellent tracking and particle-identification capabilities of the LHCb detector, and are subsequently removed by a background subtraction procedure. The $B^0_{(s)}\to \jpsi K^+\pi^-$ signal lineshape in the $m(\jpsi K^+\pi^-)$ spectrum is described by a bifurcated Gaussian core and exponential tails on both sides. A sum of two such lineshapes is used for the signal template for the mass fit, while the background lineshape is a falling exponential. The exponential tails in the signal lineshape are fixed from the simulation and all other parameters are allowed to vary in the fit, performed as a binned $\chi^2$ minimisation. An example mass fit result is given in the \ifthenelse{\boolean{isPRL}}{Supplemental material~\cite{suppl}.}{Appendix.} The cumulative signal yield in the $\mkpi\in[745,1545]$~\mev region is $554,\!500 \pm 800$.

The strategy in this analysis is to examine the hypothesis that non-exotic $K^\ast_J$ contributions alone can explain all features of the data. Under the approximation that the muon mass can be neglected and within a narrow \mkpi bin, the \CP-averaged transition matrix element squared is~\cite{spd-paper,Dey:2016oun}
\begin{align}
\label{eqn:rate_M2_XP1P2}
\!\!|\mathcal{M}|^2 &= \!\sum_{\eta}  \Big| \sum_{\lambda,J} \sqrt{2J+1} \mathcal{H}^{\eta,J}_\lambda d^J_{\lambda,0} (\thetav) d^1_{\lambda,\eta} (\thetal) e^{i\lambda \chi} \Big|^2,
\end{align}
where $\mathcal{H}^{\eta,J}_\lambda$ are the $K^\ast_J$ helicity amplitudes and $d^j_{m',m}$ are Wigner rotation matrix elements. The helicities of the outgoing lepton and $K^\ast_J$ are ${\eta= \pm 1}$ and ${\lambda\in \{0,\pm 1\}}$, respectively. Parity conservation in the electromagnetic $\jpsi \to \mun \mup$ decay leads to the relation ${\mathcal{H}^{+,J}_\lambda = \mathcal{H}^{-,J}_\lambda\equiv  \mathcal{H}^J_\lambda}$. The differential decay rate of {\BdToJPsimmKpi} with the \Kp\pim system including spin-$J$ partial waves with $J\leq J^k_{\rm max}$ can be written as 
\begin{equation}
\label{eqn:vector_moments}
\left[\frac{\deriv\Gamma^k}{\deriv\Omega^{\phantom{k}}}\right]_{J_{\rm max}^k} \propto \sum^{n_{\rm max}^k}_{i=1} f_i (\Omega) \Gamma^k_i,
\end{equation}
where the angular part in Eq.~\ref{eqn:rate_M2_XP1P2} has been expanded in an orthonormal basis of angular functions, $f_i(\Omega)$. Here, $k$ enumerates the \mkpi bin under consideration and ${\deriv\Omega = \deriv\!\ctl\,\deriv\!\ctv\,\deriv\chi}$ is the angular phase space differential element. The angular basis functions, $f_i(\Omega)$, are constructed from spherical harmonics, \mbox{$Y^m_l \equiv Y^m_l (\thetal,\chi)$}, and reduced spherical harmonics, \mbox{${P^m_l \equiv \sqrt{2 \pi}Y^m_l(\thetav,0)}$}, and are given in \ifthenelse{\boolean{isPRL}}{are given in the Supplemental material~\cite{suppl}.}{are given in the Appendix.}

The $\Gamma^k_i$ moments are observables that have an overall \mkpi dependence, but within a narrow \mkpi bin, this dependence can be neglected. The number of moments for the $k^{\rm th}$ bin, $n_{\rm max}^k$, depends on the allowed spin of the highest partial wave, $J_{\rm max}^k$, and is given by~\cite{Dey:2016oun}
\begin{equation}
\label{eqn:nmax}
n_{\rm max}^k = 28 + 12 \times (J_{\rm max}^k - 2), \text{  for } J_{\rm max}^k > 2.
\end{equation}
Thus, for spin 3 onward, each additional higher spin component leads to 12 additional moments. In contrast to previous analyses, $\deriv\!\ctl\,\deriv\chi$ is not integrated over, which would have resulted in integrating over 10 out of these 12 moments, for each additional spin. Due to the orthonormality of the $f_i (\Omega)$ basis functions, the angular observables, $\Gamma^k_i$, can be determined from the data in an unbiased fashion using a simple counting measurement~\cite{spd-paper}. For the $k^{\rm th}$ \mkpi bin, the background-subtracted raw moments are estimated as
\begin{equation}
 \label{eqn:raw_moments}
\Gamma^k_{i, {\rm raw}} =  \sum_{p=1}^{n^k_{\rm sig}} f_i(\Omega_p)  - x^k\sum_{p=1}^{n^k_{\rm bkg}} f_i(\Omega_p),
\end{equation}
where $\Omega_p$ refers to the set of angles for a given event in this \mkpi bin. The corresponding covariance matrix is
\begin{equation}
 \label{eqn:raw_covariance}
\!\! C^k_{ij, {\rm raw}} = \sum_{p=1}^{n^k_{\rm sig}} f_i(\Omega_p)f_j(\Omega_p)   + (x^k)^2\sum_{p=1}^{n^k_{\rm bkg}} f_i(\Omega_p)f_j(\Omega_p).
\end{equation}
Here, $n^k_{\rm sig}$ and $n^k_{\rm bkg}$ correspond to the number of candidates in the signal and background regions, respectively. The signal region is defined within $\pm 15$~\mev of the known \Bz mass, and the background region spans the range $\mJkpi\in[5450,5560]$~\mev. The scale factor, $x^k$, is the ratio of the estimated number of background candidates in the signal region divided by the number of candidates in the background region and is used to normalise the background subtraction.

To unfold effects from the detector efficiency including event reconstruction and selection, an efficiency matrix, $E^k_{ij}$, is used. It is obtained from simulated signal events generated according to a phase space distribution, uniform in $\Omega$, as
\begin{equation}
 \label{eqn:eff_mat}
E^k_{ij} =  \sum_{p=1}^{n^k_{\rm sim}} w^k_p f_i(\Omega_p) f_j(\Omega_p).
\end{equation}
The $w^k_p$ weight factors correct for differences between data and simulation, and the summation is over simulated and reconstructed events. They are derived using the $\Bz\to \jpsi K^\ast(892)^0$ control mode, as described in Refs.~\cite{Aaij:2015oid,Aaij:2016kqt}. The efficiency-corrected moments and covariance matrices are estimated as
\begin{align}
\Gamma^k_i &= \left( \left(E^k\right)^{-1} \right)_{il} \Gamma^k_{l, {\rm raw}},\\ 
C^k_{ij} &= \left( \left(E^k\right)^{-1} \right)_{il} C^k_{lm, {\rm raw}} \left( \left(E^k\right)^{-1} \right)_{jm}.
\label{eqn:true_moments}
\end{align}
The first moment, $ \Gamma^k_1$, corresponds to the overall rate. The remaining moments and the covariance matrix are normalised to this overall rate as $\overline{\Gamma}^k_i \equiv \Gamma^k_i / \Gamma^k_1$ and
\begin{align}
\label{eqn:norm_moments}
\overline{C}^k_{ij, {\rm stat}} &=  \left[ \frac{C^k_{ij}}{\left(\Gamma_1^k\right)^2} + \frac{\Gamma^k_i \Gamma^k_j}{\left(\Gamma^k_1\right)^4} C^k_{11} - \frac{\Gamma^k_i C^k_{1j} + \Gamma^k_j C^k_{1i}}{\Gamma^k_1  \left(\Gamma_1^k\right)^2  }\right],
\end{align}
for $i,j \in \{2,\ldots,\nmaxk\}$. 

The normalisation with respect to the total rate renders the analysis insensitive to any overall systematic effect not correlated with $\deriv \Omega$ in a given \mkpi bin. The uncertainty from limited knowledge of the background is included in the second term in Eq.~\ref{eqn:raw_covariance}. The effect on the normalised moments, $\overline{\Gamma}^k_i$, due to the uncertainty in the $x^k$ scale factors from the mass fit, is found to be negligible. The effect due to the limited simulation sample size compared to the data is small and accounted for using pseudoexperiments. The last source of systematic uncertainty is the effect of finite resolution in the reconstructed angles. The estimated biases in the measured $\overline{\Gamma}^k_i$ moments are added as additional uncertainties.

The dominant contributions to $\BdToJPsiKpi$ are from the $K^\ast(892)^0$ and $K^\ast_2(1430)^0$ states. To maximise the sensitivity to any exotic component, the dominant $K^\ast(892)^0$ region that serves as a background for any non-$K^\ast_J$ component, the analysis is performed on the ${\mkpi \in [1085,1445]}$~\mev region, as marked by the dot-dashed lines in Fig.~\ref{fig:Kst_res}. The value of $J^k_{\rm max}$ depends on \mkpi, with higher spin states suppressed at lower \mkpi values, due to the orbital angular momentum barrier factor~\cite{PhysRevD.5.624}. As seen from Fig.~\ref{fig:Kst_res}, only states with spin $J=\{0,1\}$ contribute below $\mkpi \sim 1300$~\mev and spin $J=\{0,1,2\}$ below $\mkpi\sim1600$~\mev. As a conservative choice, $J^k_{\rm max}$ is taken to be one unit larger than these expectations, 
\begin{align}
\label{eqn:kstaronly}
 \!\! \jmaxk =\begin{cases}
                  2{\rm  \;\;for\;\;} 1085 \leq \mkpi < 1265~{\rm \mev},\\
                  3{\rm  \;\;for\;\;} 1265 \leq \mkpi < 1445~{\rm \mev}.
                 \end{cases}
\end{align}

Any exotic component in the $\jpsi\pim$ or $\jpsi \Kp$ system will reflect onto the entire basis of $K^\ast_J$ partial waves and give rise to nonzero contributions from $P_l(\ctv)$ components for $l$ larger than those needed to account for $K^\ast_J$ resonances. From the completeness of the $f_i (\Omega)$ basis, a model with large enough $J^k_{\rm max}$ also describes any exotic component in the data. For a given value of $\mkpi$, there is a one-to-one correspondence between $\cos\theta_V$ and the variables $m(\jpsi\pim)$ or $m(\jpsi \Kp)$. Therefore a complete basis of $P_l(\ctv)$ partial waves also describes any arbitrary shape in $m(\jpsi\pim)$ or $m(\jpsi \Kp)$, for a given $\mkpi$ bin. The series is truncated at a value large enough to describe the relevant features of the distribution in data, but not so large that it follows bin-by-bin statistical fluctuations. A value of $J^k_{\rm max}=15$ is found to be suitable.

\begin{figure}
 \centering
 \includegraphics[width=0.95\linewidth]{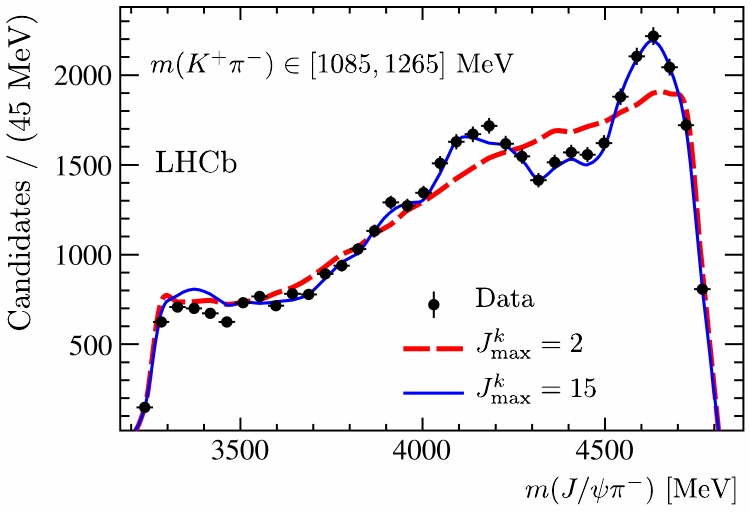}
 \caption{Comparison of $m(J\psi \pi^-)$ in the $\mkpi\in[1085,1265]$~\mev region between the background-subtracted data and simulated events weighted by moments models with $J^k_{\rm max}=2$ and $J^k_{\rm max}=15$.}
\label{fig:j15_spd_comparison_1145-1175}
\end{figure}

For the $k^{\rm th}$ \mkpi bin, the probability density function (pdf) for the $J^k_{\rm max}$ model is
\begin{align}
\label{eqn:pdf_defn}
\mathcal{P}_{\jmaxk}(\Omega) &= \displaystyle \frac{1}{\sqrt{8\pi}} \left[\frac{1}{\sqrt{8\pi}} + \sum_{i=2}^{n^k_{J_{\rm max}}} \overline{\Gamma}^k_i f_i(\Omega) \right].
\end{align}
Simulated events generated uniformly in $\Omega$, after incorporating detector efficiency effects and weighting by the pdf in Eq.~\ref{eqn:pdf_defn}, are expected to match the background-subtracted data. The background subtraction is performed using the $sPlot$ technique~\cite{Pivk:2004ty}, where the weights are determined from fits to the invariant $m(\jpsi K^+\pi^-)$ distributions described previously. Figure~\ref{fig:j15_spd_comparison_1145-1175} shows this comparison between the background-subtracted data and weighted simulated events in the $\mkpi\in[1085,1265]$~\mev region. The $J^k_{\rm max}=2$ model clearly misses the peaking structures in the data around $m(\jpsi\pi^-)=4200$~\mev and $4600$~\mev. This inability of the $J^k_{\rm max}=2$ model to describe the data, even though the first spin 2 state, $K^\ast_2(1430)^0$, lies beyond this mass region, strongly points toward the presence of exotic components. These could be four-quark bound states, meson molecules, or possibly dynamically generated features such as cusps.

To obtain a numerical estimate of the significance of exotic states, the likelihood ratio test is employed between the null hypothesis ($K^\ast_J$-only, from Eq.~\ref{eqn:kstaronly}) and the exotic hypothesis ($J^k_{\rm max}=15$) pdfs, denoted ${\cal P}^k_{K^\ast_J}$ and ${\cal P}^k_{\rm exotic}$, respectively. The test statistic used in the likelihood ratio test is defined as
\ifthenelse{\boolean{isPRL}}{\begin{widetext}}{}
\begin{align}
\label{eqn:data_chi2_test}
\ifthenelse{\boolean{isPRL}}{
\!\!\!\! \Delta (-2 \log {\cal L})\Big|_k \equiv -\displaystyle \sum_{p=1}^{n^k_{\rm sig}} 2 \log \frac{{\cal P}^k_{K^\ast_J}(\Omega_p)  }{{\cal P}^k_{\rm exotic} (\Omega_p)  }  + \displaystyle   x^k \sum_{p=1}^{n^k_{\rm bkg}} 2 \log \frac{{\cal P}^k_{K^\ast_J}(\Omega_p)  }{{\cal P}^k_{\rm exotic}(\Omega_p)} + 2\;(n^k_{\rm sig} - x^k n^k_{\rm bkg}) \log \frac{\int {\cal P}^k_{K^\ast_J}(\Omega) \epsilon(\Omega) \deriv \Omega  }{ \int {\cal P}^k_{\rm exotic}(\Omega) \epsilon(\Omega) \deriv \Omega  },}{
& \Delta (-2 \log {\cal L})\Big|_k \equiv -\displaystyle \phantom{x^k} \sum_{p=1}^{n^k_{\rm sig}} 2 \left[ \log\left( \frac{{\cal P}^k_{K^\ast_J}(\Omega_p)  }{{\cal P}^k_{\rm exotic} (\Omega_p)  } \right)  \right] + \displaystyle   x^k \sum_{p=1}^{n^k_{\rm bkg}} 2 \left[ \log \left( \frac{{\cal P}^k_{K^\ast_J}(\Omega_p)  }{{\cal P}^k_{\rm exotic}(\Omega_p)} \right)  \right] +\nonumber \\
 &  \hspace{4cm} 2 \times (n^k_{\rm sig} - x^k n^k_{\rm bkg}) \times \log\left( \frac{\int {\cal P}^k_{K^\ast_J}(\Omega) \epsilon(\Omega) \deriv \Omega  }{ \int {\cal P}^k_{\rm exotic}(\Omega) \epsilon(\Omega) \deriv \Omega  }  \right),
}
\end{align}
\ifthenelse{\boolean{isPRL}}{\end{widetext}}{}
for the $k^{\rm th}$ \mkpi bin, where $\epsilon(\Omega)$ denotes the 3-dimensional angular detector efficiency in this bin, derived from the simulation weighted to match the data in the $\Bz$ production kinematics. The last term in Eq.~\ref{eqn:data_chi2_test} ensures normalization of the relevant pdf and is calculated from simulated events that pass the reconstruction and selection criteria
\begin{align}
E^k_i &\equiv \displaystyle \sum_{p=1}^{n^k_{\rm sim}} w^k_p f_i(\Omega_p), \\
\int {\cal P}_{\jmaxk}(\Omega) \epsilon(\Omega) \deriv \Omega &\propto \displaystyle \sum_{i=1}^{\nmaxk} \Gamma^k_i E^k_i.
\end{align}
Results from individual \mkpi bins are combined to give the final test statistic ${\Delta (-2 \log {\cal L}) = \displaystyle \sum_k  \Delta (-2 \log {\cal L})\Big|_k}$.

From Eq.~\ref{eqn:nmax} the number of degrees-of-freedom (ndf) increases by 12 for each additional spin-$J$ wave in each \mkpi bin. From Eq.~\ref{eqn:kstaronly}, for the $J^k_{\rm max}=2$ and $3$ choices, $\Delta{\rm ndf}= 12\times (15-2)=156$ and $12\times(15-3)=144$, respectively, between the exotic and $K^\ast_J$-only pdf's for each \mkpi bin. Each additional degree-of-freedom between the exotic and $K^\ast_J$-only pdf adds approximately one unit to the computed $\Delta (-2 \log {\cal L})$ in the data due to increased sensitivity to the statistical fluctuations, and $\Delta (-2 \log {\cal L})$ is therefore not expected to be zero even if there is no exotic contribution in the data. The expected $\Delta (-2 \log {\cal L})$ distribution in the absence of exotic activity is evaluated using a large number of pseudoexperiments. For each \mkpi bin, 11,000 pseudoexperiments are generated according to the $K^\ast_J$-only model with the moments varied according to the covariance matrix. The number of signal and background events for each pseudoexperiment are taken to be those  measured in the data. The detector efficiency obtained from simulation is parameterised in 4D. Each pseudoexperiment is analyzed in exactly the same way as the data, where an independent efficiency matrix is generated for each pseudoexperiment. This accounts for the limited sample size of the simulation for the efficiency unfolding. The pseudoexperiments therefore represent the data faithfully at every step of the processing.

\begin{figure}
 \centering
 \includegraphics[width=\linewidth]{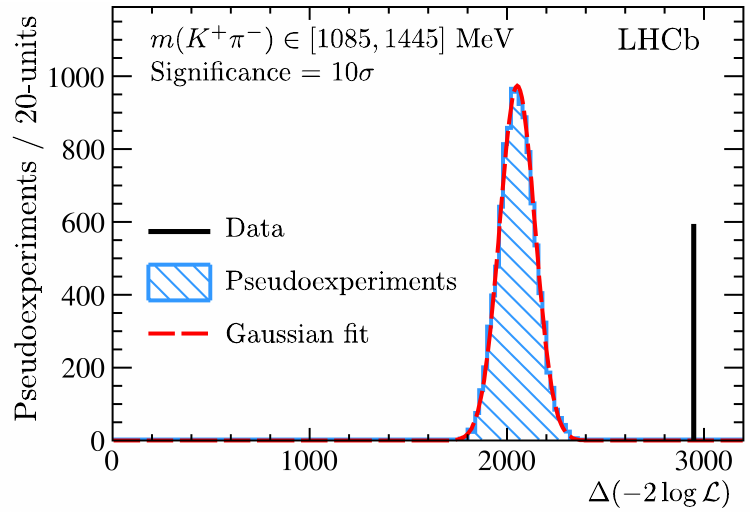}
 \caption{Likelihood-ratio test for exotic significance. The data shows a $10\sigma$ deviation from the pseudoexperiments generated according to the null hypothesis ($K^\ast_J$-only contributions).}
\label{fig:DLL_1085_1445}
\end{figure}

Figure~\ref{fig:DLL_1085_1445} shows the distribution of $\Delta (-2 \log {\cal L})$ from the pseudoexperiments in the $\mkpi\in[1085,1445]$~\mev region comprising six \mkpi bins each with the $J^k_{\rm max}=2$ or $3$ choice. A fit to a Gaussian profile gives $\Delta (-2 \log {\cal L})\approx 2051$ between the null and exotic hypothesis, even in the absence of any exotic contributions. This value is consistent with the na{\"\i}ve expectation $\Delta({\rm ndf})= 1800$ from the counting discussed earlier. The value of $\Delta (-2 \log {\cal L})$ for the data, as marked by the vertical line in Fig.~\ref{fig:DLL_1085_1445}, shows a deviation of more than $10\sigma$ from the null hypothesis, corresponding to the distribution of the pseudoexperiments. The uncertainty due to the quality of the Gaussian profile fit in Fig.~\ref{fig:DLL_1085_1445} is found to be negligible. The choice of large $J^k_{\rm max}$ for ${\cal P}^k_{\rm exotic}$, as well as the detector efficiency and calibration of the simulation, are systematically varied in pseudoexperiments, with significance for exotic components in excess of $6\sigma$ observed in each case.

\begin{figure}
 \centering
 \includegraphics[width=1.05\linewidth]{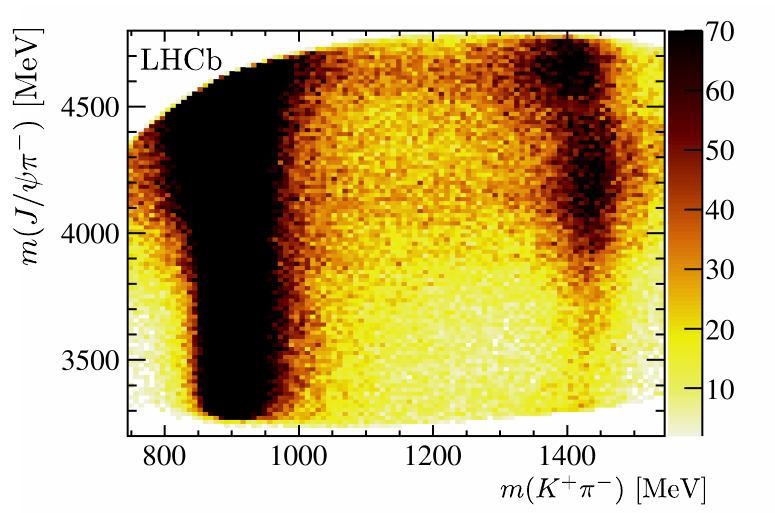}
 \caption{Background-subtracted 2D distribution of $m(\jpsi \pi^-)$ versus $\mkpi$ in the region ${\mkpi\in[745,1545]}$~\mev. The intensity ($z$-axis) scale has been highly truncated to limit the strong $K^\ast(892)^0$ contribution.}
\label{fig:dalitz}
\end{figure}

In summary, employing the Run~1 LHCb dataset, non-$K^\ast_J$ contributions in ${B^0\to \jpsi K^+\pi^-}$ are observed with overwhelming significance. Compared to the previous BaBar analysis~\cite{Aubert:2008aa} of the same channel, the current study benefits from a 40-fold increase in signal yield and a full angular analysis of the decay topology. The method relies on a novel orthonormal angular moments expansion and, aside from a conservative limit on the highest allowed $K^\ast_J$ spin for a given $\mkpi$ invariant mass, makes no other assumption about the $K^+\pi^-$ system. Figure~\ref{fig:dalitz} shows a scatter plot of $m(\jpsi \pi^-)$ against $\mkpi$ in the background-subtracted data. While the model-independent analysis performed here does not identify the origin of the non-$K^\ast_J$ contributions, structures are visible at ${m(\jpsi \pi^-)\approx 4200}$~\mev, close to the exotic state reported previously by Belle~\cite{Chilikin:2014bkk}, and at $m(\jpsi\pi^-)\approx 4600$~\mev. To interpret these structures as exotic tetraquark resonances and measure their properties will require a future model-dependent amplitude analysis of the data.

\section*{Acknowledgements}
%
%
\noindent We express our gratitude to our colleagues in the CERN
accelerator departments for the excellent performance of the LHC. We
thank the technical and administrative staff at the LHCb
institutes.
We acknowledge support from CERN and from the national agencies:
CAPES, CNPq, FAPERJ and FINEP (Brazil); 
MOST and NSFC (China); 
CNRS/IN2P3 (France); 
BMBF, DFG and MPG (Germany); 
INFN (Italy); 
NWO (Netherlands); 
MNiSW and NCN (Poland); 
MEN/IFA (Romania); 
MSHE (Russia); 
MinECo (Spain); 
SNSF and SER (Switzerland); 
NASU (Ukraine); 
STFC (United Kingdom); 
NSF (USA).
We acknowledge the computing resources that are provided by CERN, IN2P3
(France), KIT and DESY (Germany), INFN (Italy), SURF (Netherlands),
PIC (Spain), GridPP (United Kingdom), RRCKI and Yandex
LLC (Russia), CSCS (Switzerland), IFIN-HH (Romania), CBPF (Brazil),
PL-GRID (Poland) and OSC (USA).
We are indebted to the communities behind the multiple open-source
software packages on which we depend.
Individual groups or members have received support from
AvH Foundation (Germany);
EPLANET, Marie Sk\l{}odowska-Curie Actions and ERC (European Union);
ANR, Labex P2IO and OCEVU, and R\'{e}gion Auvergne-Rh\^{o}ne-Alpes (France);
Key Research Program of Frontier Sciences of CAS, CAS PIFI, and the Thousand Talents Program (China);
RFBR, RSF and Yandex LLC (Russia);
GVA, XuntaGal and GENCAT (Spain);
the Royal Society
and the Leverhulme Trust (United Kingdom);
Laboratory Directed Research and Development program of LANL (USA).




\clearpage


\appendix


{\noindent\normalfont\bfseries\Large Appendix}\\

\vspace{0.5cm}

{\noindent\normalfont\bfseries\large A. Angle conventions}\\
\begin{figure*}[h]
\centering
\subfigure[]{
\centering
\includegraphics[width=0.45\linewidth]{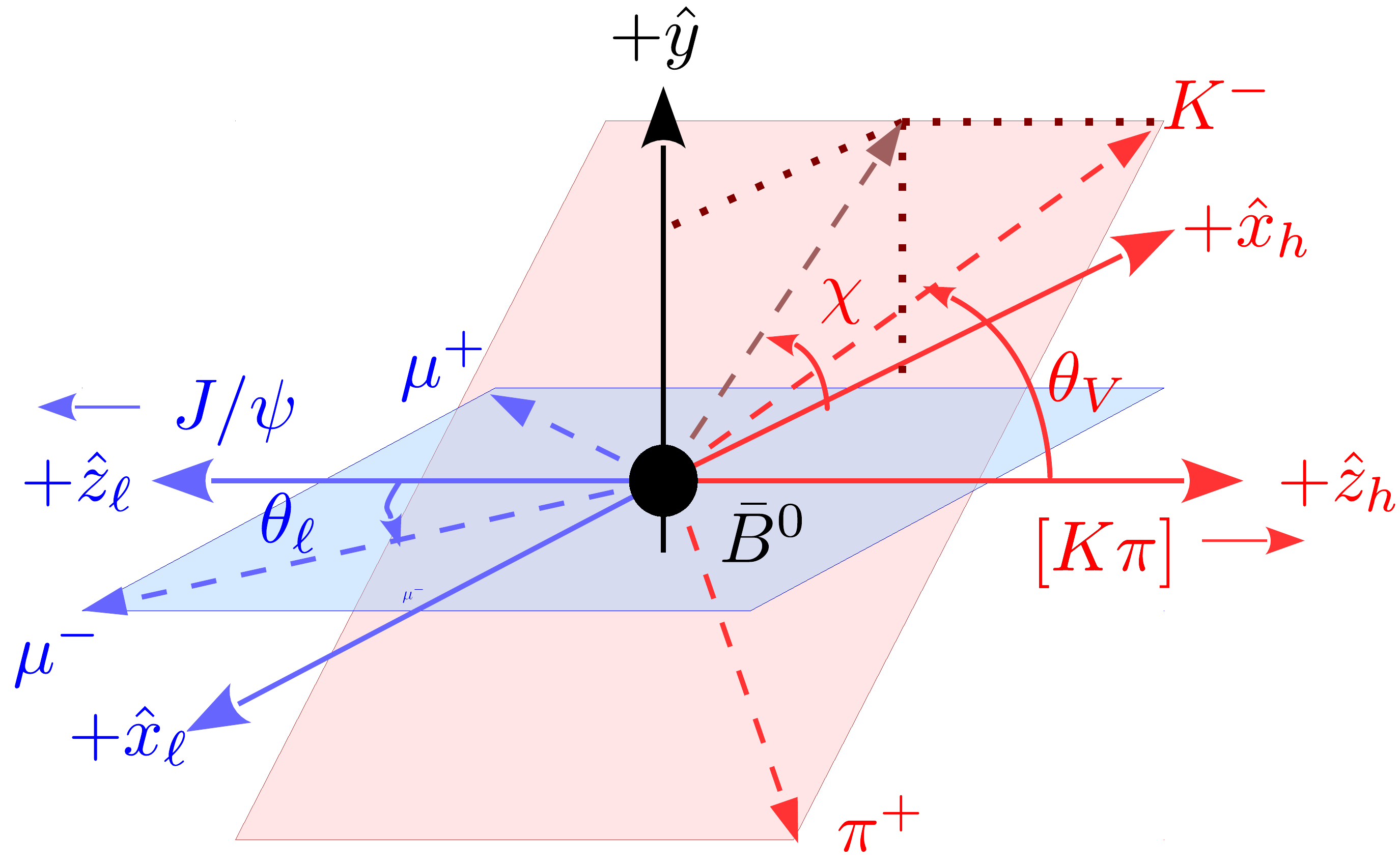}
}
\subfigure[]{
\centering
\includegraphics[width=0.45\linewidth]{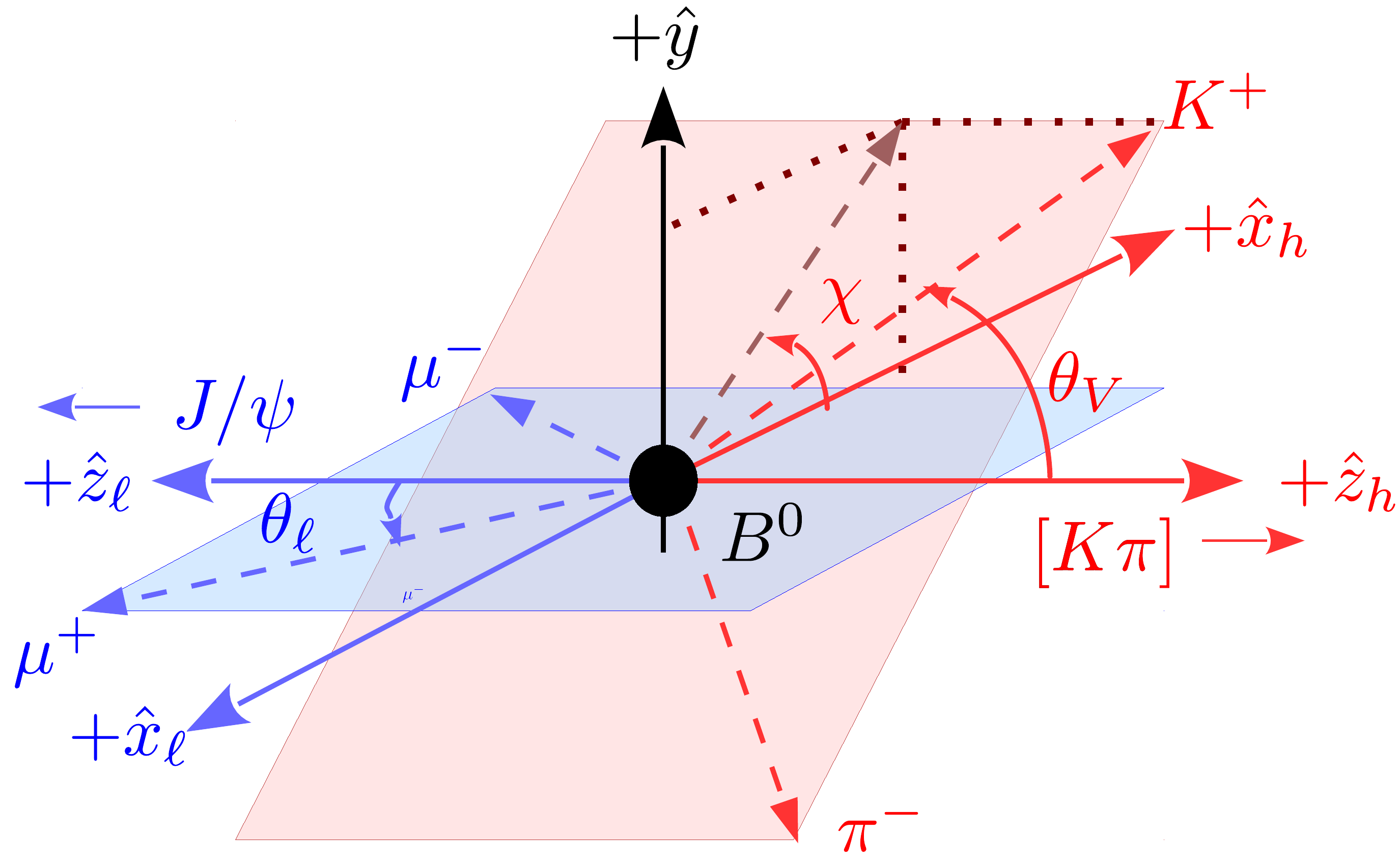}
}
\caption{Angle conventions as described in Ref.~\cite{spd-paper} for (a) $\Bzb \to \jpsi (\to \mun \mup) \Km \pip$ and (b) ${\Bz \to \jpsi (\to \mup \mun) \Kp \pim}$. The leptonic ($\mup\mun$) and hadronic ($K^+\pi^-$) frames are back-to-back with a common $\hat{y}$ axis.}
\label{fig:angle_conventions}
\end{figure*}

\noindent
The four kinematic variables for the process $\Bz \to \jpsi (\to \mup \mun) \Kp \pim$ are the invariant mass \mkpi, and the three angles $\{\thetal,\thetav,\chi\}$. The angle conventions for $\Bz$ and $\Bzb$ are depicted in Fig.~\ref{fig:angle_conventions}. Assuming negligible direct \CP violation and production asymmetry, the rate expression remains the same between the charge-conjugate modes.\\

\vspace{0.5cm}

{\noindent\normalfont\bfseries\large B. Example mass fit result in a particular bin}\\

\begin{figure}[h]
 \centering
 \includegraphics[width=\linewidth]{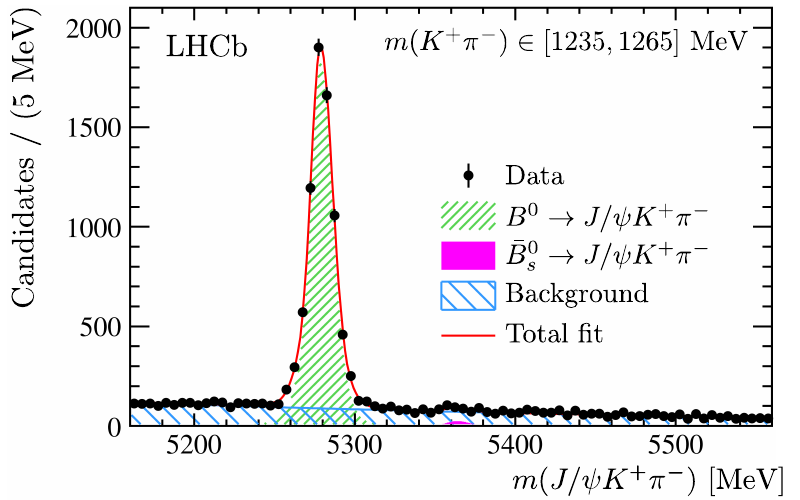}
 \caption{Fit to the invariant mass $m(\jpsi K^+\pi^-)$ in the $\mkpi \in [1235,1265]$~\mev bin.}
\label{fig:massfit_body}
\end{figure}

\noindent Figure~\ref{fig:massfit_body} shows an example mass fit result for the $\mkpi \in [1235,1265]$~\mev bin.

\vspace{0.5cm}

{\noindent\normalfont\bfseries\large C. Further comparison between the $\bm{J^k_{\rm max}=2}$ and $\bm{J^k_{\rm max}=15}$ models}\\

\begin{figure}[h]
 \centering
 \includegraphics[width=\linewidth]{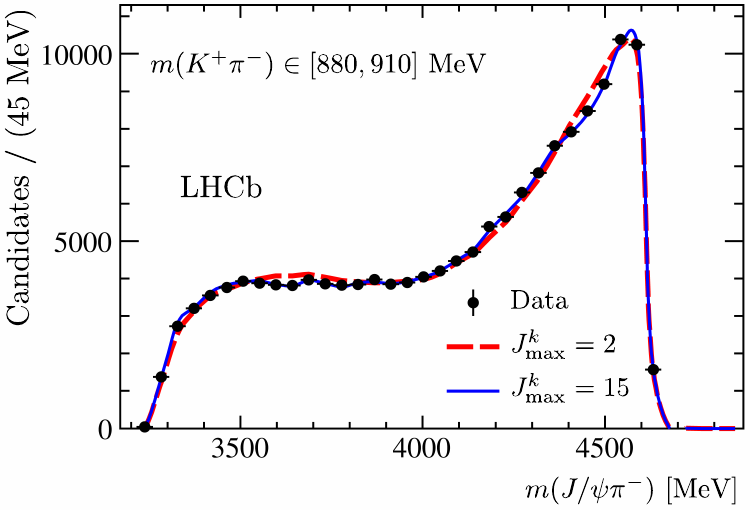}
 \caption{Comparison of $m(J/\psi\pi^-)$ in the $m(K^+\pi^-)\in[880,910]$~MeV region between the background-subtracted data and simulation data weighted by moments models with $J^k_{\rm max}=2$ and $J^k_{\rm max}=15$.}
\label{fig:j15_spd_comparison_880-910}
\end{figure}

\noindent Figure~\ref{fig:j15_spd_comparison_880-910} shows a comparison between the $J^k_{\rm max}=2$ and $J^k_{\rm max}=15$ moments models in the $m(K^+\pi^-)=895\pm15$~MeV bin. Since the spin-1 $K^\ast(892)^0$ resonance strongly dominates here, the two models are compatible.\\

\begin{figure}[h]
 \centering
 \includegraphics[width=\linewidth]{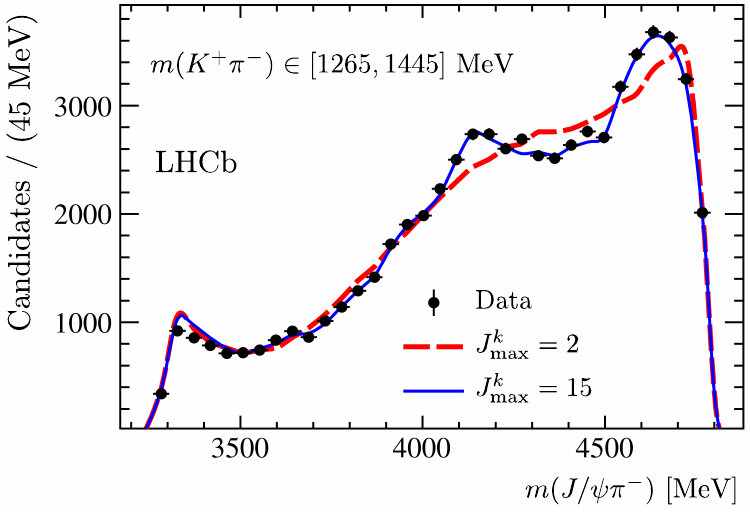}
 \caption{Comparison of $m(J/\psi\pi^-)$ in the $m(K^+\pi^-)\in[1265,1445]$~MeV region between the background-subtracted data and simulation data weighted by moments models with $J^k_{\rm max}=2$ and $J^k_{\rm max}=15$.}
\label{fig:j15_spd_comparison_1265-1445}
\end{figure}

\noindent Figure~\ref{fig:j15_spd_comparison_1265-1445} shows a comparison between the $J^k_{\rm max}=2$ and $J^k_{\rm max}=15$ moments models in the $\mkpi \in [1265,1445]$~\mev.\\

\vspace{0.5cm}

{\noindent\normalfont\bfseries\large D. Angular moments definitions}\\

\noindent The transversity basis amplitudes, $\mathcal{H}^J_{\{\parallel, \perp\}}$, are defined as
\begin{align}
\mathcal{H}^J_\pm &= (\mathcal{H}^J_\parallel\pm \mathcal{H}^J_\perp)/\sqrt{2}
\end{align}
and the amplitudes for spin $J\in \{0,1,2\}$ are denoted as $S$, $H_{\{0,\parallel,\perp\}}$ and $D_{\{0,\parallel,\perp\}}$, respectively. For $K^\ast_J$ contributions up to $J=2$, there are 28 angular moments from the expansion of Eq.~\ref{eqn:rate_M2_XP1P2}, as explicitly listed in Table~\ref{tab:spd_mom} in terms of the transversity amplitudes. The addition of $K^\ast_J$ states from spin-3 onward results in 12 moments for each additional spin. The form of the moments are listed in Table~\ref{tab:angular_terms_j_gt_2}, leading to the expression appearing in Eq.~\ref{eqn:nmax} of the main text. Further details can be obtained from Refs.~\cite{spd-paper,Dey:2016oun}.\\

\begin{table}
\caption{The transversity-basis moments of the 28 orthonormal angular functions $f_i(\Omega)$ in Eq.~\ref{eqn:vector_moments} till spin-2 in the \kpi system.}
\label{tab:spd_mom}
\centering
\resizebox{\textwidth}{!}{
\begin{tabular}{c|c|c}
 $i$    &   $f_i(\Omega)$             & $\Gamma^{{\rm tr}}_i(\qsq)$  \\ \hline \hline
 1   &   $P^0_0 Y^0_0$     &  $\left[ \hzsq + \hpasq + \hpesq + \ssq + \dzsq + \dpasq + \dpesq\right]$ \\ \hline
 2   &   $P^0_1 Y^0_0$     &  $2\left[\frac{2}{\sqrt{5}} \rhzdz + \rshz + \sqrt{\frac{3}{5}}  \rel( H_\parallel D^{\ast}_\parallel + H_\perp D^{\ast}_\perp  )\right]$  \\ \hline
 3   &   $P^0_2 Y^0_0$     &  $\frac{\sqrt{5}}{7}$ (\dpasq + \dpesq) - $\frac{1}{\sqrt{5}}$ (\hpasq + \hpesq) + $\frac{2}{\sqrt{5}}$ \hzsq  + $\frac{10}{7\sqrt{5}}$ \dzsq + $2$ \rsdz \\  \hline
 4   &   $P^0_3 Y^0_0$     &  $\frac{6}{\sqrt{35}} \left[ - \rel(H_\parallel D^{\ast}_\parallel +  H_\perp D^{\ast}_\perp)  + \sqrt{3} \rhzdz  \right]$ \\  \hline
 5   &   $P^0_4 Y^0_0$     &  $\frac{2}{7} \left[ -2 (\dpasq + \dpesq) + 3 \dzsq \right] $ \\  \hline
 6   &   $P^0_0 Y^0_2$     &  $\frac{1}{2 \sqrt{5}} \left[ (\dpasq + \dpesq) + (\hpasq + \hpesq) - 2 \ssq - 2 \dzsq - 2 \hzsq \right]$ \\  \hline
 7   &   $P^0_1 Y^0_2$     &  $\left[ \frac{\sqrt{3}}{5} \rel(H_\parallel D^{\ast}_\parallel  + H_\perp D^{\ast}_\perp) - \frac{2}{\sqrt{5}} \rel(S H^{\ast}_0)  - \frac{4}{5} \rel(H_0 D^{\ast}_0)\right] $   \\ \hline
 8   &   $P^0_2 Y^0_2$     &  $ \left[ \frac{1}{14} (\dpasq + \dpesq) - \frac{2}{7} \dzsq - \frac{1}{10} (\hpasq + \hpesq) - \frac{2}{5} \hzsq - \frac{2}{\sqrt{5}} \rsdz \right]$   \\  \hline
 9   &   $P^0_3 Y^0_2$     &  $ - \frac{3}{5 \sqrt{7}} \left[ \rel( H_\parallel D^{ \ast}_\parallel + H_\perp D^{ \ast}_\perp) + 2 \sqrt{3} \rel(H_0 D^{ \ast}_0 ) \right] $ \\  \hline
 10  &   $P^0_4 Y^0_2$     &  $ -\frac{2}{7 \sqrt{5}}  \left[ \dpasq + \dpesq + 3 \dzsq \right] $   \\  \hline
 11  &   $P^1_1 \sqrt{2}\rel(Y^1_2)$ &  $-\frac{3}{\sqrt{10}} \left[ \sqrt{\frac{2}{3}} \rel(H_\parallel S^{ \ast}) - \sqrt{\frac{2}{15}} \rel(H_\parallel D^{ \ast}_0  ) + \sqrt{\frac{2}{5}} \rel(D_\parallel H^{ \ast}_0 ) \right] $  \\  \hline
 12  &   $P^1_2 \sqrt{2}\rel(Y^1_2)$ &  $-\frac{3}{5} \left[ \rel( H_\parallel H^{ \ast}_0)  + \sqrt{\frac{5}{3}} \rel (D_\parallel S^{ \ast})  + \frac{5}{7 \sqrt{3}} \rel(D_\parallel D^{\ast}_0) \right] $  \\  \hline
 13  &   $P^1_3 \sqrt{2}\rel(Y^1_2)$ &  $-\frac{6}{5 \sqrt{14}} \left[2 \rel(D_\parallel H^{\ast}_0)  + \sqrt{3} \rel(H_\parallel D^{\ast}_0 ) \right] $  \\  \hline
 14  &   $P^1_4 \sqrt{2}\rel(Y^1_2)$ &  $- \frac{6}{7\sqrt{2}} \rel(D_\parallel D^{\ast}_0)$   \\  \hline
 15  &   $P^1_1 \sqrt{2}\img(Y^1_2)$ &  $3 \left[ \frac{1}{\sqrt{15}} \img(H_\perp S^{\ast}) + \frac{1}{5} \img(D_\perp H^{\ast}_0) - \frac{1}{5 \sqrt{3}}  \img(H_\perp D^{\ast}_0) \right]  $   \\  \hline
 16  &   $P^1_2 \sqrt{2}\img(Y^1_2)$ &  $ 3\left[ \frac{1}{7 \sqrt{3}} \img(D_\perp D^{\ast}_0)  + \frac{1}{5} \img(H_\perp H^{\ast}_0)  + \frac{1}{\sqrt{15}} \img(D_\perp S^{\ast})   \right] $   \\  \hline
 17  &   $P^1_3 \sqrt{2}\img(Y^1_2)$ &  $\frac{6}{5 \sqrt{14}} \left[ 2 \img(D_\perp H^{\ast}_0)  + \sqrt{3} \img(H_\perp D^{\ast}_0) \right]  $    \\  \hline
 18  &   $P^1_4 \sqrt{2}\img(Y^1_2)$ &  $\frac{6}{7\sqrt{2}} \img(D_\perp D^{\ast}_0)$   \\  \hline
 19  &   $P^0_0 \sqrt{2}\rel(Y^2_2)$ &  $-\frac{3}{2\sqrt{15}} \left[ (\hpasq - \hpesq) + (\dpasq - \dpesq) \right] $   \\  \hline
 20  &   $P^0_1 \sqrt{2}\rel(Y^2_2)$ &  $-\frac{3}{5} \left[ \rel(H_\parallel D^{\ast}_\parallel)   - \rel(D_\perp H^{\ast}_\perp) \right] $   \\  \hline
 21  &   $P^0_2 \sqrt{2}\rel(Y^2_2)$ &  $\frac{\sqrt{3}}{2} \left[ - \frac{1}{7} (\dpasq - \dpesq)   + \frac{1}{5} ( \hpasq - \hpesq ) \right] $   \\  \hline
 22  &   $P^0_3 \sqrt{2}\rel(Y^2_2)$ &  $\frac{3}{5} \sqrt{ \frac{3}{7}} \left[ \rel(H_\parallel D^{\ast}_\parallel)   - \rel(D_\perp H^{\ast}_\perp) \right] $   \\  \hline
 23  &   $P^0_4 \sqrt{2}\rel(Y^2_2)$ &  $\frac{2}{7} \sqrt{ \frac{3}{5}}  (\dpasq - \dpesq) $  \\  \hline
 24  &   $P^0_0 \sqrt{2}\img(Y^2_2)$ &  $\sqrt{\frac{3}{5}} \left[ \img(H_\perp H^{\ast}_\parallel) + \img(D_\perp D^{\ast}_\parallel) \right] $    \\  \hline
 25  &   $P^0_1 \sqrt{2}\img(Y^2_2)$ &  $\frac{3}{5} \img(  H_\perp D^{\ast}_\parallel +  D_\perp H^{\ast}_\parallel )  $   \\ \hline
 26  &   $P^0_2 \sqrt{2}\img(Y^2_2)$ &  $ \sqrt{3} \left[\frac{1}{7} \img(D_\perp D^{\ast}_\parallel)   - \frac{1}{5} \img(H_\perp H^{\ast}_\parallel)\right] $   \\ \hline
 27  &   $P^0_3 \sqrt{2}\img(Y^2_2)$ &  $-\frac{3}{5} \sqrt{ \frac{3}{7}}  \img(D_\perp H^{\ast}_\parallel + H_\perp D^{\ast}_\parallel)  $   \\ \hline
 28  &   $P^0_4 \sqrt{2}\img(Y^2_2)$ &  $-\frac{4}{7} \sqrt{\frac{3}{5}}  \img(D_\perp D^{\ast}_\parallel) $
\end{tabular}
}
\end{table}

\begin{table}
\caption{The 12 angular terms for each additional spin-$J$ wave in the \kpi system, for $J\geq3$.}
\label{tab:angular_terms_j_gt_2}
\centering
\begin{tabular}{c|c}
$i$ & $f_i(\Omega)$\\
\hline \hline
$1$ & $P^0_{2J-1} Y^0_0$\\ \hline
$2$ & $P^0_{2J\phantom{-1}} Y^0_0$ \\ \hline \hline
$3$ & $P^0_{2J-1} Y^0_2$ \\ \hline
$4$ & $P^0_{2J\phantom{-1}} Y^0_2$  \\ \hline
$5$ & $P^1_{2J-1} \sqrt{2} Re(Y^1_2)$  \\ \hline
$6$ & $P^1_{2J\phantom{-1}} \sqrt{2} Re(Y^1_2)$ \\ \hline
$7$ & $P^1_{2J-1}\sqrt{2} Im(Y^1_2)$  \\ \hline
$8$ & $P^1_{2J\phantom{-1}} \sqrt{2} Im(Y^1_2)$  \\ \hline
$9$ & $P^0_{2J-1}\sqrt{2} Re(Y^2_2)$  \\ \hline
$10$ & $P^0_{2J\phantom{-1}} \sqrt{2} Re(Y^2_2)$  \\ \hline
$11$ & $P^0_{2J-1} \sqrt{2} Im(Y^2_2)$  \\ \hline
$12$ & $P^0_{2J\phantom{-1}}\sqrt{2} Im(Y^2_2)$
\end{tabular}
\end{table}

\clearpage


\addcontentsline{toc}{section}{References}
\setboolean{inbibliography}{true}
\bibliographystyle{LHCb}
\bibliography{main}

\newpage
\centerline
{\large\bf LHCb Collaboration}
\begin
{flushleft}
\small
R.~Aaij$^{29}$,
C.~Abell{\'a}n~Beteta$^{46}$,
B.~Adeva$^{43}$,
M.~Adinolfi$^{50}$,
C.A.~Aidala$^{77}$,
Z.~Ajaltouni$^{7}$,
S.~Akar$^{61}$,
P.~Albicocco$^{20}$,
J.~Albrecht$^{12}$,
F.~Alessio$^{44}$
% ...

{\footnotesize \it

% ...
$ ^{7}$Universit{\'e} Clermont Auvergne, CNRS/IN2P3, LPC, Clermont-Ferrand, France\\
% ...
$ ^{12}$Fakult{\"a}t Physik, Technische Universit{\"a}t Dortmund, Dortmund, Germany\\
% ...
$ ^{20}$INFN Laboratori Nazionali di Frascati, Frascati, Italy\\
% ...
$ ^{29}$Nikhef National Institute for Subatomic Physics, Amsterdam, Netherlands\\
% ...
$ ^{43}$Instituto Galego de F{\'\i}sica de Altas Enerx{\'\i}as (IGFAE), Universidade de Santiago de Compostela, Santiago de Compostela, Spain\\
$ ^{44}$European Organization for Nuclear Research (CERN), Geneva, Switzerland\\
% ...
$ ^{46}$Physik-Institut, Universit{\"a}t Z{\"u}rich, Z{\"u}rich, Switzerland\\
% ...
$ ^{50}$H.H. Wills Physics Laboratory, University of Bristol, Bristol, United Kingdom\\
% ...
$ ^{61}$University of Cincinnati, Cincinnati, OH, United States\\
% ...
$ ^{63}$Syracuse University, Syracuse, NY, United States\\
% ...
$ ^{77}$University of Michigan, Ann Arbor, United States, associated to $^{63}$\\
% ...
}
\end{flushleft}

\end{document}